\documentstyle[preprint,prc,aps]{revtex}
\tightenlines
\begin{document}
\title{HOW THE ``H-PARTICLE" UNRAVELS THE QUARK DYNAMICS}
\author{Fl. Stancu and S. Pepin}
\address{Universit\'{e} de Li\`ege, Institut de Physique B.5, Sart Tilman,
B-4000 Li\`ege 1, Belgium}
\author{L. Ya. Glozman}
\address{Institute for Theoretical Physics, University of Graz, 8010 Graz,
Austria}
\date{\today}
\maketitle

\begin{abstract}
It is shown that the short-range part of the Goldstone boson exchange
interaction between the constituent quarks which explains baryon
spectroscopy and the short-range repulsion in the NN system induces a strong
short-range repulsion in the flavour-singlet state of the $\Lambda\Lambda$
system with $J^P=0^+$. It then suggests that a deeply bound H-particle should
not exist. We also compare our approach with other models employing different
hyperfine interactions between quarks in the nonperturbative regime of QCD.
\end{abstract}

\vspace{1cm}

{\bf I.} Soon after the suggestion that the hyperfine 
splitting in hadrons should be
due to the colour-magnetic interaction between quarks
\cite{RGG75,CJJ74} it has been noted by Jaffe \cite{Ja77} that the dibaryon
$uuddss$ with $J^P=0^+$, $I=0$, called the H particle, is stable against strong
decays. Its mass turned out to be about 80 MeV below the $\Lambda\Lambda$
threshold. The reason is that in this case the flavour-singlet state in the 6q
system is allowed and the colour-magnetic interaction gives more attraction
for the most favourable configuration $[33]_{CS}$ than for two well-separated
$\Lambda$-hyperons. In Jaffe's picture the H-particle should be a compact
object in contrast to the molecular-type structure of the deuteron.

Since Jaffe's prediction many calculations in a variety of models have
appeared \cite{all}. They give a wide range of predicted masses depending
on the model. Realistic calculations usually predict a well-bound H-particle.
In particular, the quark-cluster calculations suggest that the implications of
the colour-magnetic interaction are radically different in two-nucleon and
in coupled YY-YN systems. While in the former case the colour-magnetic
interaction between quarks gives rise to a strong short-range repulsion in the
NN system, in the latter, there appears either a soft 
attraction or a soft repulsion
at short-range \cite{OSY87} when the linear combination of the coupled
channels is close to a flavour-singlet state. This soft short-range
interaction, reinforced by the medium and long-range attraction coming from
the meson exchange between lambdas, provides a bound state with a binding
energy of the order 10-20 MeV
\cite{SZBFK88} or even 60-120 MeV \cite{KSY90}. However, a simple
quark-cluster variational basis, used in these calculations, 
is rather poor at short-range. While it is not
so important for the baryon-baryon systems with strong repulsion at
short-range, this shortcoming becomes crucial for the
$\Lambda\Lambda - N\Xi - \Sigma\Sigma$ system, with the quantum numbers of H.
As soon as a simple quark-cluster variational basis is properly extended, a
very deeply bound state with the binding energy of about 250 MeV 
is  found \cite{WM97}.

The existence or non-existence of the H-particle has to be settled by
experiment. Since approximately 20 years several experiments have been set
for ``hunting" the H-particle. The very recent high-sensitivity search at
Brookhaven \cite{sto97} gives no evidence for the production of deeply bound
H, the production cross section being one order of magnitude below theoretical
estimates.

It has recently been suggested that in the low-energy regime, light and
strange baryons should be considered as systems of three constituent quarks
with a $QQ$ interaction ($Q$ is a constituent quark, 
to be contrasted with a current
quark $q$) that is formed of a central confining part and a chiral interaction
that is mediated by Goldstone bosons between constituent quarks \cite{GR96}.
Indeed, at low temperatures and densities, the underlying chiral symmetry of
QCD is spontaneously broken by the QCD vacuum. This implies that the valence
quarks acquire a constituent (dynamical) mass, which is related to the
quark condensates $<\bar{q}q>$ and at the same time the Goldstone bosons
$\pi, K, \eta$ appear, which couple directly to the constituent quarks
\cite{We79}. It has been shown that the hyperfine splittings as well as the
correct ordering of positive and negative parity states in spectra of baryons
with valence $u,d,s$ quarks are produced in fact, not by the 
colour-magnetic part
of the one-gluon exchange interaction (OGE), but by the short-range part of the
Goldstone boson exchange (GBE) interaction \cite{GR96,G97,semrel}. This
short-range part of the GBE interaction has just opposite sign as compared to
the Yukawa potential tail and is much stronger at short interquark
separations. There is practically no room for OGE in light baryon spectroscopy
and any appreciable amount of colour-magnetic interaction, in addition to GBE,
destroys the spectrum
\cite{semrel2}. The same short-range part of the GBE interaction, which
produces good baryon spectra, also induces a short-range repulsion in the NN
system
\cite{SPG97}. Thus it is interesting to study the short-range interaction in
the
$\Lambda\Lambda$ system and the stability of the H-particle in the GBE model.\\

 {\bf II.} For a qualitative insight it is convenient first to 
consider a schematic
quark-quark interaction which neglects the radial dependence of the GBE
interaction. In this model, the short-range part of the GBE interaction between
the constituent quarks is approximated by:
\begin{equation}
 V_{\chi} = - C_{\chi} \sum_{i<j}  \lambda_{i}^{F} . \lambda_{j}^{F}
\vec{\sigma}_i . \vec{\sigma}_j ,
\label{opFS}
\end{equation}
where $\lambda_{i}^{F}$ with an implied summation over F (F=1,2,...,8) are
the quark flavour Gell-Mann matrices and $\vec{\sigma}$ the spin matrices.
The minus sign of the interaction (\ref{opFS}) is related to the sign of the
short-range part of the GBE interaction (which is opposite to that of the
Yukawa potential tail), crucial for the hyperfine splittings in baryon
spectroscopy.

In a harmonic oscillator basis, $\hbar \omega$ and the constant $C_{\chi}$ 
implied by the schematic model (\ref{opFS}), can
be determined from $\Delta-N$ and $N(1440)-N$ splittings to be $C_{\chi} =
29.3$ MeV, $\hbar \omega \sim 250$ MeV \cite{GR96}.

The colour- and flavour-singlet $uuddss$ states are described by $[222]_C$ and
$[222]_F$ Young diagrams respectively. For the S-wave relative motion of two
$s^3$ clusters, the spatial symmetries of the $6Q$ 
system are $[6]_O$ and $[42]_O$
and for the spin S=0 the corresponding spin symmetry is $[33]_S$. The
antisymmetry condition requires
$[f]_{FS} = [\tilde{f}]_{OC}$, where $[\tilde{f}]$ is the conjugate of $[f]$.
Thus, among the states given by the inner products:
\begin{eqnarray}
[33]_S \times [222]_F & = & [33]_{FS} + [411]_{FS} + [2211]_{FS} +
[1^6]_{FS}, \\
\, [6]_O \times [222]_C & = & [222]_{OC}, \\
 \, [42]_O \times [222]_C & = & [42]_{OC} + [321]_{OC} + [222]_{OC} +
[3111]_{OC} + [21111]_{OC} 
\end{eqnarray}
only the four states are allowed:
\begin{equation}
\begin{array}{ccc}
|1> &=& |[6]_O [33]_{FS} [222]_{OC}> \\
|2> &=& |[42]_O [33]_{FS} [222]_{OC}> \\
|3> &=& |[42]_O [411]_{FS} [3111]_{OC}> \\
|4> &=& |[42]_O [2211]_{FS} [42]_{OC}> \\
\end{array}
\label{basis}
\end{equation}
For these states the expectation values of the interaction (\ref{opFS}) 
can be easily calculated in terms of the Casimir operators eigenvalues for
the groups $SU(6)_{FS}, SU(3)_F$ and $SU(2)_S$ using the formula given in
Appendix A of Ref.\cite{SPG97}. The corresponding matrix elements are given in
Table 1. Thus the interaction (\ref{opFS}) is attractive for the states
$|1>-|3>$ and repulsive for $|4>$. This suggests that it is
a good approximation to restrict the basis to $|1>, |2>$ and $|3>$ for
the diagonalization of a more realistic Hamiltonian. Keeping in mind that the
spatial symmetry
$[6]_O$ is compatible with the lowest non-excited $s^6$ configuration, one can
roughly evaluate the energy of the lowest state relative to
the $2\Lambda$ threshold as:
\begin{eqnarray}
<s^6 [6]_O [33]_{FS} | H_0 + V_{\chi} | s^6 [6]_O [33]_{FS}> -  2<s^3 [3]_O
[3]_{FS} | H_0 + V_{\chi} | s^3 [3]_O [3]_{FS}> \nonumber \\
= 4 C_{\chi} + 3/4 \hbar \omega = 305 \mbox{ MeV}, 
\end{eqnarray}
where $H_0$ is the kinetic energy in the $6Q$ system.
While here and below we use notations of the shell model, it is always
assumed that the center of mass motion is removed. In deriving 
the kinetic energy, $3/4
\hbar \omega$, we have neglected the mass difference between u,d and s
constituent quarks. The pair-wise colour electric confinement contribution is
exactly the same for $s^6$ configuration and for two well separated $s^3$
clusters, so it cancels out.

This simple estimate shows that the lowest ``compact" flavour-singlet $6Q$ 
state
with quantum  numbers $J^P=0^+, I=0, S=-2$ lies a few hundreds MeV above the
$\Lambda\Lambda$ threshold.\\ 

{\bf III.} In a more quantitative calculation we use 
the Hamiltonian \cite{GR96,G97}:
\begin{equation}
H= \sum_{i=1}^6 m_i + \sum_i \frac{\vec{p}_{i}^{2}}{2m} - \frac {(\sum_i
\vec{p}_{i})^2}{12m} + \sum_{i<j} V_{conf}(r_{ij}) + \sum_{i<j} V_\chi(
\vec{r_{ij}})
\label{ham}
\end{equation}
where the confining interaction is:
\begin{equation}
 V_{conf}(r_{ij}) = -\frac{3}{8}\lambda_{i}^{c}\cdot\lambda_{j}^{c} \, C
\, r_{ij}
\label{conf}
\end{equation}
and the spin-spin component of the GBE interaction between the constituent
quarks i and j reads:
\begin{eqnarray}
V_\chi(\vec r_{ij})
&=&
\left\{\sum_{F=1}^3 V_{\pi}(\vec r_{ij}) \lambda_i^F \lambda_j^F \right.
\nonumber \\
&+& \left. \sum_{F=4}^7 V_{\rm K}(\vec r_{ij}) \lambda_i^F \lambda_j^F
+V_{\eta}(\vec r_{ij}) \lambda_i^8 \lambda_j^8
+V_{\eta^{\prime}}(\vec r_{ij}) \lambda_i^0 \lambda_j^0\right\}
\vec\sigma_i\cdot\vec\sigma_j,
\label{VCHI}
\end{eqnarray}

\noindent
where $\lambda^0=\sqrt{2/3}~{\bf 1}$ ($\bf 1$ is the $3 \times 3$ unit matrix).
The interaction (\ref{VCHI}) includes $\pi, K, \eta$ and $\eta'$ exchanges. In
the large-$N_c$ limit, when axial anomaly vanishes \cite{W79}, the spontaneous
breaking of the chiral symmetry $U(3)_L \times U(3)_R \rightarrow U(3)_V$
implies a ninth Goldstone boson \cite{CW80}, which corresponds to the flavour
singlet $\eta'$. Under real conditions, for $N_c=3$, a certain contribution
from the flavour singlet remains and the $\eta'$ must thus be included in the
GBE interaction.

 In the simplest case, when both the constituent quarks and mesons are
point-like particles and the boson field satisfies
the linear Klein-Gordon equation, one has the following spatial dependence
for the meson-exchange potentials \cite{GR96} :
 
\begin{equation}V_\gamma (\vec r_{ij})=
\frac{g_\gamma^2}{4\pi}\frac{1}{3}\frac{1}{4m^2}
\{\mu_\gamma^2\frac{e^{-\mu_\gamma r_{ij}}}{ r_{ij}}-4\pi\delta (\vec r_{ij})\}
,  \hspace{5mm} (\gamma = \pi, K, \eta, \eta' )
\label{POINT} \end{equation}

\noindent
where $\mu_\gamma$ are the  meson masses and $g_\gamma^2/4\pi$ are
the quark-meson coupling constants given below.
 
Eq. (\ref{POINT}) contains both the traditional long-range
Yukawa potential as well as a
$\delta$-function term. It is the latter  that is of crucial importance
for baryon spectroscopy and short-range $NN$ interaction since it has a proper
sign to provide the correct hyperfine splittings in baryons and is becoming
highly dominant at short range.
Since one deals with structured particles (both the constituent quarks and
pseudoscalar mesons) of finite extension, one must
smear out the $\delta$-function in (\ref{POINT}).
 In Ref. \cite{GPP96} a smooth Gaussian term has been employed instead of
the $\delta$-function

\begin{equation}4\pi \delta(\vec r_{ij}) \Rightarrow \frac {4}{\sqrt {\pi}}
\alpha^3 \exp(-\alpha^2(r-r_0)^2), \label{CONTACT} \end{equation}
 
\noindent
where $\alpha$ and $r_0$ are adjustable parameters.

The parameters of the Hamiltonian (\ref{ham})-(\ref{VCHI}) are \cite{GPP96}:

$$\frac{g_{\pi q}^2}{4\pi} = \frac{g_{\eta q}^2}{4\pi} = 0.67,\,\,
\frac{g_{\eta ' q}^2}{4\pi} = 1.206,$$
$$r_0 = 0.43 \, {\rm fm}, ~\alpha = 2.91 \, {\rm fm}^{-1},~~
 C= 0.474 \, {\rm fm}^{-2}, $$
\begin{equation}
 \mu_{\pi} = 139 \, {\rm MeV},~ \mu_{\eta} = 547 \, {\rm MeV},~
\mu_{\eta'} = 958 \, {\rm MeV}.
\label{PAR} \end{equation}
 
\noindent
The Hamiltonian (\ref{ham})-(\ref{PAR}) with constituent
masses
$m_{u,d}=340$ MeV, $m_s=440$ MeV provides a very satisfactory description of
the low-lying N and $\Delta$ spectra in a fully dynamical nonrelativistic
3-body calculation \cite{GPP96} as well as of the strange baryon spectra
\cite{GPPVW96}. However, this parametrization should be considered as an
effective one only. Indeed, the volume integral of GBE interaction should
be zero \cite{semrel}, while in the parametrization above this is not so
because of the off-shift $r_0$ of the "contact" term. This problem has
been overcome in a semirelativistic parametrization \cite{semrel}, where
the parameters of the potential as well as the form of the contact term
are very different. However, at present a semirelativistic description
of baryons cannot be applied to a study of baryon-baryon interactions
since such a study can be done only nonrelativistically
with $s^3$ wave functions for ground state baryons.
Thus one needs an effective nonrelativistic parametrization of the $QQ$
potential which would provide correct energies of octet-decuplet
baryons with $s^3$ ansatz for their wave functions. 
The nonrelativistic parametrization
above meets this requirement. In particular,
$<\Lambda | H | \Lambda >$ takes its minimal values of 1165.4
MeV at a harmonic oscillator parameter value of $\beta=0.449$ fm, i.e. only
about 40 MeV above the actual value, obtained in the dynamical 3-body
calculations \cite{GPPVW96}.
 Since in this paper we study only qualitative
effects, related to the spin-flavour structure and sign of the short-range
part of the GBE interaction, we consider such an approach
as a reasonable framework. 

 We calculate the
 potential in the flavor-singlet $S=-2$ two-baryon system at zero separation
between clusters in the  adiabatic (Born-Oppenheimer) approximation
defined as:
\begin{equation}
V(R) = <H>_R - <H>_{\infty},
\label{born}
\end{equation}
where $R$ is a collective coordinate which is the separation distance
between the two  $s^3$ clusters, $<H>_R$ is
the lowest expectation value of the Hamiltonian describing the $6Q$ system
at fixed $R$ and $<H>_{\infty} = 2 m_\Lambda$,
i.e. the energy of two well separated lambdas, obtained with the same
Hamiltonian.

It has been shown by Harvey \cite{H81} that when the separation $R$ between
two $s^3$ clusters approaches 0,  then only two types
of  $6Q$ configurations survive: $|s^6 [6]_O>$ and
$|s^4p^2 [42]_O>$. 
Thus in order to extract an effective  potential
at zero separation between clusters in the adiabatic 
approximation we diagonalize the Hamiltonian 
 (\ref{ham})-(\ref{PAR}) in the basis
of the first three states defined by (\ref{basis}). All the necessary
matrix elements are calculated with the help of the fractional
parentage technique, also used
in a study of the short-range $NN$ interaction in  ref. \cite{SPG97}.

 We find the lowest eigenvalue of the flavour-singlet state 
$J^P=0^+$ to be 847 MeV above the $\Lambda\Lambda$ threshold.
According to  (\ref{born}), 
there is a strong short-range repulsion
in a two-baryon flavor-singlet $S=-2$ system in $^1S_0$ wave.
 It then
definitely suggests that within the physical picture under discussion the
compact (well bound) H-particle should not exist.

The value of the repulsion given above depends on the way the kinetic energy
of the $6Q$ system  was calculated. For simplicity,
 we considered in the kinetic
energy term only, that u,d and s quarks have the same mass $\bar{m}=(4 m_u +2
m_s)/6$. We have also carried calculations in the extreme limits $\bar{m}=m_u$
and
$\bar{m}=m_s$ and obtained 1050 MeV and 531 MeV respectively, which is the
energy above the
$\Lambda\Lambda$ threshold. This extreme values just prove that the strong
repulsion persists in any case.\\ 

{\bf IV.} This result is in a sharp contrast with  models based on the 
colour-magnetic interaction as the hyperfine interaction between quarks,
which gives a short-range attraction in  a two-baryon flavor-singlet
 $S=-2$ system in the $^1S_0$ wave.
We consider this result as an additional evidence in favour of the GBE model to
be the dominant hyperfine interaction between the constituent quarks. Indeed,
a deeply bound H-particle is definitely excluded by experiment \cite{sto97}
and the colour-magnetic interaction, at variance with GBE, implies such a
deeply bound state  (see I).

There are suggestions that the instanton-induced ('t Hooft) interaction could
be important for the hyperfine splittings in baryons \cite{SR89}. Assuming that
the instanton-induced interaction in $QQ$ pairs 
is responsible for the essential
part of the $\Delta-N$ hyperfine splitting, the deeply bound H-particle should
also disappear \cite{TO91}. This interaction is very strong and attractive
in colour-singlet $q\bar{q}$ pseudoscalar channel and could be indeed
responsible for the chiral symmetry spontaneous breaking in QCD vacuum
\cite{CDG78} and be the most important interaction in mesons. Thus to the 
extent the 't Hooft interaction contributes to  $q\bar{q}$ pseudoscalar
pairs, it is automatically taken into account when one includes the GBE
interaction  in $QQ$ pairs (the 't Hooft interaction could be 
responsible, at least in part, for the pole in t-channel). However, the 
"direct" 't Hooft interaction in $qq$ pairs is rather weak.
There are also indications from lattice QCD that the "direct"
instanton-induced interaction in $qq$ pairs cannot be responsible 
for the $\Delta-N$ splitting.
For example, the
$\Delta-N$ splitting disappears after cooling \cite{CGHN94} (only instantons
survive the cooling procedure), while it is appreciable before cooling. There
is also  evidence from lattice QCD that the hyperfine splittings are
related mostly to $q\bar{q}$ excitations in baryons, but not to forces mediated
by gluonic fields in $qq$ pairs \cite{LD94}. There are also simple symmetry
arguments showing that "direct" 't Hooft interaction in $QQ$ pairs 
cannot provide correct ordering of
lowest positive and negative parity states in light and strange baryon spectra
\cite{GR96} (for baryon spectra obtained in such a model in a
nonperturbative calculation see second paper in Ref.\cite{SR89}. 
From the results obtained in this paper, one can see, indeed, that the lowest
positive parity excitations in all parts of the spectrum
-N(1440),$\Delta$(1600),$\Lambda$(1600),$\Sigma$(1660),...- lie much above
the negative parity excitations).

One should also mention the QCD sum rule estimate for the H-particle
\cite{KOH94}. It is shown there that there is no qualitative difference
between the two-nucleon system and the two-lambda one (including the flavor
singlet channel), which strongly supports our point of view.\\

{\bf V.} Here we have considered the $6Q$ 
$S=-2$ system in a flavour singlet state 
only ("H-particle" channel) 
and found that there appears a strong short-range repulsion in $^1S_0$
partial wave. This strong short-range repulsion implies that a deeply bound
(on nuclear scale) H-particle should not exist. The
same analysis can be extended to the $\Lambda\Lambda$ system in all allowed
flavour states.
Then, similarly to the NN system \cite{SPG97} there will appear a strong
short-range repulsion coming from the same short-range part 
of the GBE interaction.
There is however an attraction in the
$\Lambda\Lambda$ system at medium- and long-range, coming from the Yukawa 
potential tail of the GBE interaction as well as from correlated
two-pseudoscalar-meson exchange. At the moment, one cannot exclude that this
interaction could very weakly bind
$\Lambda\Lambda$ in a molecule-like system of nuclear nature. 
However, this attraction should be
similar in its origin
to the attraction in the $^1S_0$ partial wave of the NN system, which is too
weak to bind the system. A firm prediction about the existence or
non-existence of a  weakly bound
$\Lambda\Lambda$ system of nuclear nature can only be made in a fully
dynamical calculation.

\begin{table}
\renewcommand{\arraystretch}{1.5}
\caption[Expectation values]{\label{expectation} Expectation values of the
operator (\ref{opFS}) in $C_{\chi}$ units corresponding to the states
(\ref{basis}).}
\begin{tabular}{|c|c|} 
$[f]_O [f]_{FS} [f]_{OC}$ \hspace{1cm} & $<V_{\chi}>/C_{\chi}$ \hspace{5cm} \\
\tableline
$[6]_O [33]_{FS} [222]_{OC}$ \hspace{1cm}& -24 \hspace{5cm} \\
$[42]_O [411]_{FS} [3111]_{OC}$ \hspace{1cm}& -24 \hspace{5cm} \\
$[42]_O [33]_{FS} [222]_{OC}$ \hspace{1cm}& -24 \hspace{5cm} \\
$[42]_O [2211]_{FS} [42]_{OC}$ \hspace{1cm}& 8 \hspace{5cm} \\
\end{tabular}
\end{table}

\end{document}